\documentclass[11pt,twoside]{book} 
\usepackage{asp2008n}
\usepackage{times}
\usepackage{lscape}
\usepackage{epsf}

\usepackage{graphicx}
\usepackage{amssymb}
\usepackage{longtable}
\usepackage[figuresright]{rotating}
\usepackage{multirow}

\newcommand{\simless}{\mathbin{\lower 3pt\hbox {$\rlap{\raise 5pt\hbox{$\char'074$}}\mathchar"7218$}}}

\mainmatter

\newlength{\deftabcolsep}
\begin{document}


\title{The LkH$\alpha$ 101 Cluster}

\author{Sean M. Andrews and Scott J. Wolk}

\affil{Harvard-Smithsonian Center for Astrophysics \\
60 Garden Street, Cambridge, MA, 02138, USA}

\begin{abstract}
In the infrared, the heavily reddened LkH$\alpha$ 101 is one of the brightest
young stars in the sky.  Situated just north of the Taurus-Auriga complex in
the L1482 dark cloud, it appears to be an early B-type star that has been
serendipitously exposed during a rarely observed stage of early evolution,
revealing a remarkable spectrum and a directly-imaged circumstellar disk.
While detailed studies of this star and its circumstellar environment have
become increasingly sophisticated in the 50 years since \citet{herbig56} first
pointed it out, the true nature of the object still remains a mystery.  Recent
work has renewed focus on the young cluster of stars surrounding LkH$\alpha$
101, and what it can tell us about the enigmatic source at its center (e.g.,
massive star formation timescales, clustered formation mechanisms).  This
latter effort certainly deserves more intensive study.  We describe the current
knowledge of this region and point out interesting work that could be done in
the future.
\end{abstract}

\section{Introduction}

In a generalized sense, there are three distinct types of young star clusters:
($a$) high-mass star-forming regions with an associated extensive network of
low-mass stars (e.g., Orion); ($b$) quiescent environments that host low-mass
star formation exclusively (e.g., Taurus-Auriga); and ($c$) smaller clusters of
low-mass stars surrounding one or a few A/B stars.  Naturally, there is a
continuum of such types, and the picture is not quite so simple.  However, an
important goal in this line of research is to generally understand the
differences and commonalities between these cluster types in an effort to
better explain the various clustered modes of star formation and their
consequences.  The LkH$\alpha$ 101 cluster is an interesting example of the
($c$) type; a handful of B stars and a hundred or more low-mass stars with a
dominant source (LkH$\alpha$ 101) at the center.  The remarkable central source
and apparent young age for the cluster indicate that we have been afforded a
fortuitous opportunity to investigate this formation mode at a very early
time.  In this chapter, we highlight various studies of the LkH$\alpha$ 101
region, separated into sections focused on the local interstellar medium
(Sect.~2), distance estimates (Sect.~3), the embedded young cluster (Sect.~4),
and LkH$\alpha$ 101 itself (Sect.~5).  We conclude with a brief preview of a
new, comprehensive multiwavelength study of the region, and summarize the
information with an eye toward future studies (Sect.~6).

\begin{figure}[ht!]
\begin{center}
\includegraphics[height=4.7in,draft=False]{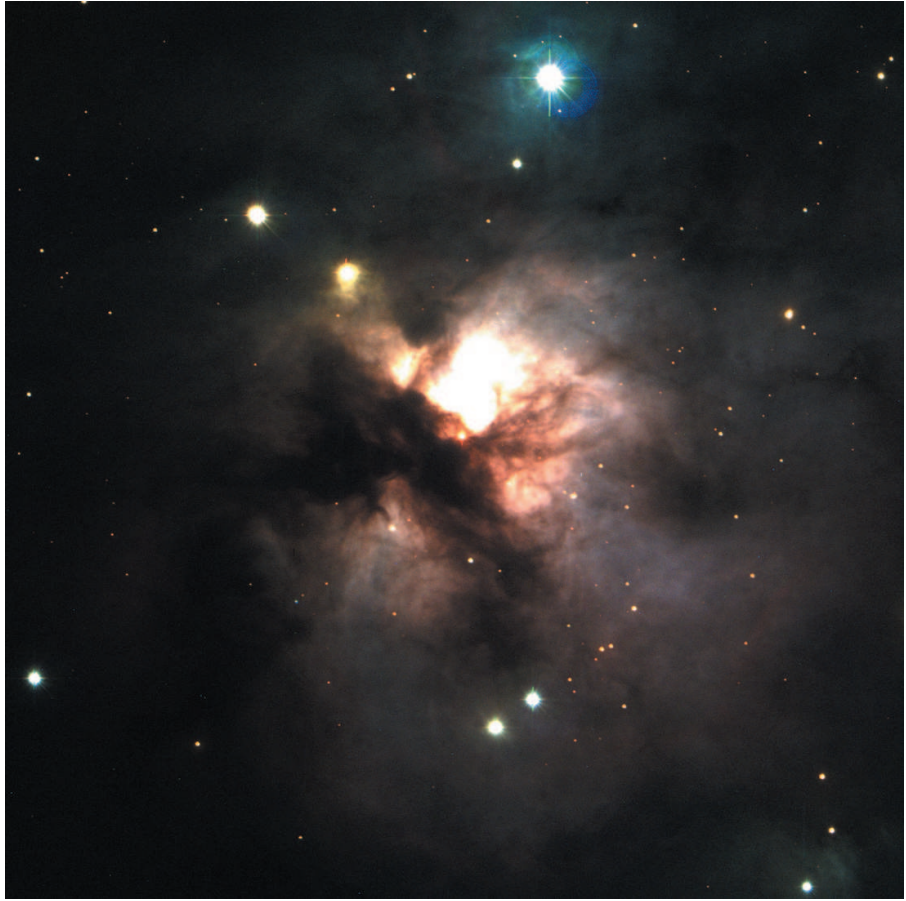}
\end{center}
\caption{Optical $VRI$ image of the NGC 1579 reflection nebula that hosts a
young cluster surrounding LkH$\alpha$ 101 \citep[from][]{herbig04}.  The image
is roughly 7 arcminutes on a side, with north up and east to the left.}
\end{figure}

\section{The Interstellar Material}

LkH$\alpha$ 101 is located just north of the Taurus-Auriga complex ($\alpha =
4^{{\rm h}}30^{{\rm m}}14.4^{{\rm s}}$, $\delta =
+35^{\circ}16\arcmin24\arcsec$ [J2000]; $l = 165.3^{\circ}$, $b =
-9.0^{\circ}$) in the L1482 dark cloud.  As Figure 1 demonstrates, the
examination of an optical image near LkH$\alpha$ 101 reveals a complex local
interstellar environment.  The most prominent feature is a dark lane which cuts
across the southeastern corner of the reddened reflection nebula NGC 1579
(discovered by William Herschel in 1788), illuminated by the apparently faint
red source LkH$\alpha$ 101 near the image center.  \citet{redman86} argue that
this dark lane is in the foreground and probably not associated with
LkH$\alpha$ 101.  That argument is further supported by the interstellar
H$_3^+$ chemistry constraints in the vicinity of an intense radiation source
\citep{brittain04}.  The reflection nebula was [mis]identified as the H\,II
region S222 because of its redness in the Palomar survey plates
\citep{sharpless59}.  However, subsequent observations showed that the
polarization pattern from dust scattering was consistent with a reflection
nebula entirely illuminated by LkH$\alpha$ 101 \citep{redman86}.  A direct
comparison of the optical spectrum of the nebula and LkH$\alpha$ 101 confirms
this conclusion \citep*{herbig04}.  The evidence for an actual H\,II region is
only inferred from the radio continuum spectrum
\citep*[e.g.,][]{brown76,dewdney86,becker88}: none of the optical or infrared
lines typical of such physical conditions are present in the spectra of the
nebula or LkH$\alpha$ 101 itself \citep{herbig04}.  The latter probably could
be explained if circumstellar material at high densities collisionally
deexcites the standard forbidden emission lines.

Additional dark clouds to the north, south, and southwest were noted in
coarse-resolution CO surveys of the region \citep{knapp76,redman86,barsony90}.
The northern cloud is visible as a lip of material at the edge of the NGC 1579
nebula (see Fig.~1).  \citet{redman86} present a schematic diagram of these
various interstellar components (see their Fig.~8), and \citet{herbig04}
discuss in some detail the small-scale structures in the medium immediately
surrounding LkH$\alpha$ 101.  Star counts and multicolor photometry
\citep*{barsony91} show that extinction is higher to the east of LkH$\alpha$
101.  In order of decreasing proximity to LkH$\alpha$ 101, the basic
interstellar environment consists of: a dense circumstellar disk/envelope; a
small H\,II region; a reflection nebula (NGC 1579); an H\,I envelope
\citep{dewdney82}; and a dark cloud (L1482) within a molecular filament.

LkH$\alpha$ 101 and its associated young cluster are embedded in this cloud
filament, denoted TGU 1096 by \citet{dobashi05} and shown in Fig.~2, that
extends northwest of the Taurus-Auriga complex ($d \approx 140$\,pc; see the
chapter in this volume by Kenyon, Gomez, \& Whitney), and overlaps in
projection with the more distant Per OB2 association ($d \approx 300$\,pc; see
the chapter in this volume by Bally et al.).  Despite the apparent proximity of
these two star-forming regions, the CO velocity of the filament is
significantly different ($V_{{\rm LSR}} = -1$\,km s$^{-1}$) than for Tau-Aur
(+6\,km s$^{-1}$) and Per OB2 \citep[+6-10\,km s$^{-1}$;][]{ungerechts87}.
Clearly this filamentary cloud and its contents are kinematically distinct from
the Tau-Aur and Per OB2 clouds.  Moreover, \citet{herbig04} note that the
interstellar Na\,I absorption lines toward two stars in the young cluster are
double, with core velocities consistent with those of Per OB2 and material in
the filamentary cloud.  As noted by those authors, this suggests that the
interstellar material, embedded cluster, and LkH$\alpha$ 101 lie \emph{beyond}
the Per OB2 complex.

\begin{figure}[ht!]
\begin{center}
\includegraphics[width=\textwidth,draft=False]{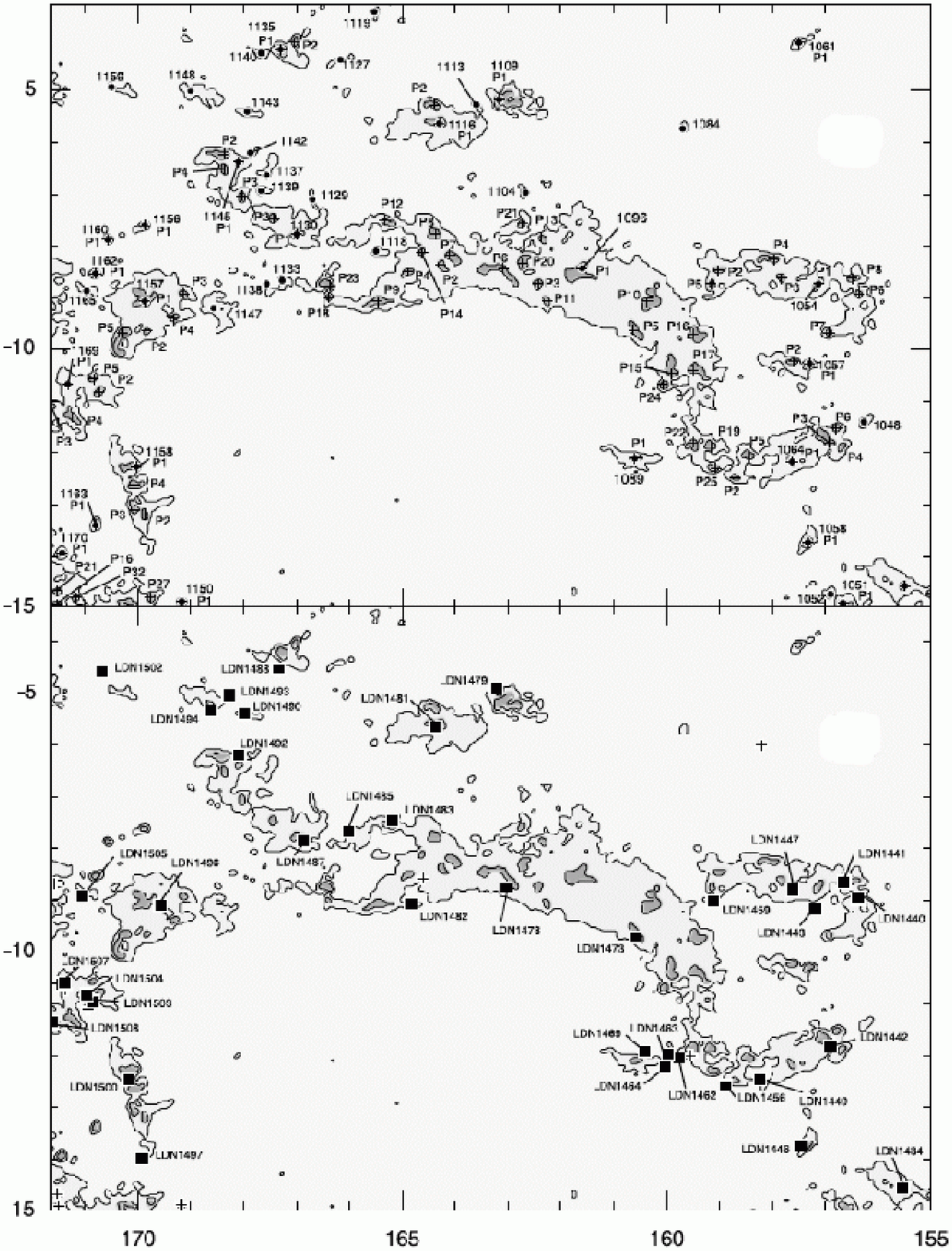}
\end{center}
\caption{Extinction map of the large-scale region (TGU 1096) surrounding the
L1482 dark cloud and the young cluster surrounding LkH$\alpha$ 101 embedded in
it.  The axes are galactic coordinates, and LkH$\alpha$ 101 is located at $l =
165.3^{\circ}$, $b = -9.0^{\circ}$.}
\end{figure}

\section{Distance Estimates}

The sky-projected proximity of LkH$\alpha$ 101 to both the Tau-Aur and Perseus
regions might be expected to create some confusion in estimating the distance
to this star and its associated young cluster.  \citet{herbig71} originally
estimated $d \approx 800$\,pc based on spectra and $UBV$ photometry of 2
early-type (B) stars near LkH$\alpha$ 101 that are associated with nebulosity.
As discussed in Sect.~2, the interstellar medium signatures also appear
consistent with a distance beyond the Perseus clouds \citep[$d >
350$\,pc;][]{ungerechts87,herbig04}.  However, \citet{stine98} argued that the
radio luminosities of some cluster stars would be an order of magnitude larger
than the mean for weak-lined T Tauri stars in Tau-Aur if the distance was as
large as 800\,pc.  Given the location on the sky and this apparent radio
luminosity discrepancy, Stine \& O'Neal suggested a much smaller $d \approx
160$\,pc.  Using high resolution infrared measurements, \citet{tuthill02}
identified both a companion star and a circumstellar disk around LkH$\alpha$
101 (see Sect.~5).  Given the proper motion of the companion and some model
constraints on the star+disk mass, they find that $d \approx 200$-500\,pc can
best explain the data, with a favored value $d \approx 340$\,pc.  Most
recently, \citet{herbig04} extended the spectroscopic parallax measurements to
40 young cluster stars with a wide range of spectral types (from mid-M to early
B) to estimate a larger distance, $d \approx 700$\,pc.

Obviously, a definitive conclusion has yet to be reached.  In the meantime, it
would pay to consider the various pieces of observational evidence in a
self-consistent manner.  Perhaps the most straight-forward distance estimate
comes from the spectroscopic parallax determinations of cluster members.  These
measurements converge on a mean distance of 700\,pc; and, although there is a
scatter around this value of approximately $\pm 200$\,pc, there are no
individual measurements consistent with a distance as low as 160\,pc.  We can
turn to the H-R diagram for a consistency check on these values using the
derived cluster age (see Sect.~4).  Interstellar reddening uncertainties will
not significantly affect the age determination because the reddening vector is
roughly parallel to the isochrones.  Therefore, the adopted distance acts to
set the cluster age.  As discussed further in Sect.~4, the cluster age is
$\sim$1\,Myr for $d = 700$\,pc.  However, if the true distance were as close as
$d = 160$\,pc, the inferred cluster age would be $\ge 10$\,Myr, and therefore
inconsistent with all the obvious indicators of youth that have been observed
in the cluster (e.g., strong H$\alpha$ emission, infrared excesses, etc.).

Along with the above age-consistency argument and the evidence from the
interstellar signatures (Sect.~2), \citet{herbig04} cautioned against adopting
the $d \approx 160$\,pc estimate advocated by \citet{stine98} because only one
of the four radio sources used by those authors is clearly associated with a
weak-line T Tauri star.  Further coupling this with the fact that none of the
other 13 weak-line T Tauri stars in the LkH$\alpha$ 101 cluster were detected
in the radio continuum, a comparison of the cluster radio luminosities with the
Tau-Aur weak-line T Tauri star mean luminosity does not present a convincing
argument for a small cluster distance.  Instead, the preponderance of evidence
suggests a large cluster distance, with most observational constraints in
agreement with $d \approx 500$-700\,pc.

\section{The Embedded Young Cluster}

The first hints of an embedded young star cluster in this region came
serendipitously from a radio study of the LkH$\alpha$ 101 stellar wind
\citep{becker88}.  The radio map revealed a ``necklace" of faint point sources
surrounding the central star that in some cases appeared to be associated with
optically detected low-mass stars.  Those initial observations were revisited
by \citet{stine98}, who identified more than a dozen compact sources, some of
them exhibiting flaring gyrosynchrotron emission similar to those seen around
weak-line T Tauri stars in Tau-Aur \citep[e.g.,][]{chiang96}.

Detailed multiwavelength observing campaigns were conducted soon after the
radio discovery \citep{barsony90,barsony91}, including broadband optical and
infrared imaging, millimeter spectral line maps, and millimeter interferometry
of the LkH$\alpha$ 101 circumstellar environment.  Those studies first claimed
a large infrared clustering of stars near LkH$\alpha$ 101, with an apparent age
gradient indicating that the central star was quite young ($\sim$10$^5$\,yr).
The latter conclusion remains somewhat an open question, due to the bright and
spatially variable nebulosity in the vicinity of the hot star.  These initial
near-infrared images were then supplemented with $L$-band photometry to better
determine the circumstellar properties of the stars in the cluster
\citep{aspin94}.  The color-color analysis in that study suggested that
$\sim$30\%\ of the surveyed stars had excess thermal emission from the inner
regions of circumstellar dust disks.  This $L$-band excess fraction would be
low for the apparent young age of the cluster (see below), according to the
fairly well-established correlation noted by \citet{haisch01} and others.
However, the completeness limit of these observations is difficult to estimate
due to the bright infrared nebulosity in the region.

\begin{figure}[ht!]
\centering
\includegraphics[height=3.25in,draft=False]{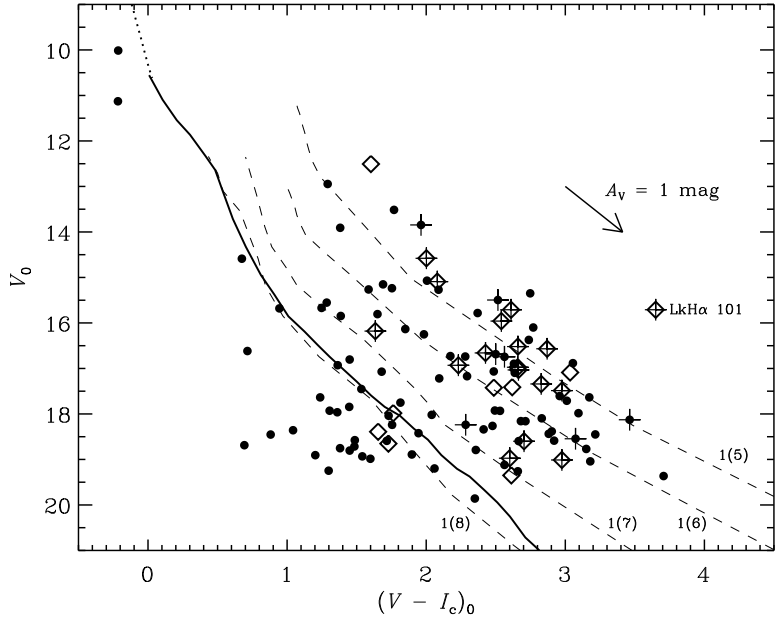}
\caption{Color-magnitude diagram adapted from \citet{herbig04}.  The sources
with known spectral types are marked with diamonds, and classical T Tauri stars
with crosses.  The solid line is the Pleiades main-sequence at $d = 700$\,pc.
The arrow shows the shift expected for an additional 1 magnitude of visual
extinction.  The dashed lines are theoretical isochrones from
\citet{dantona97}; ages are marked near the bottom of the figure.}
\end{figure}

More recently, \citet{herbig04} presented a comprehensive look at the embedded
cluster and LkH$\alpha$ 101 itself using optical and infrared imaging and
spectroscopy.  Deep $BVRI$ imaging (see Fig.~1) enabled these authors to
perform a standard analysis of the H-R diagram in an attempt to determine the
cluster age.  Figure 3 shows their reddening-corrected $V$, $V-I$
color-magnitude diagram for the cluster, along with some representative
theoretical pre$-$main-sequence isochrones \citep{dantona97}.  Supplementary
spectroscopic data revealed 35 H$\alpha$ emission line stars (excluding
LkH$\alpha$ 101) scattered around the cluster.  Thirteen ($\sim$40\%) of these
H$\alpha$ emission line stars have equivalent widths less than 10\,\AA\ (i.e.,
are weak-line T Tauri stars).  The identifications of \citet{herbig04},
celestial coordinates, representative optical and near-infrared magnitudes, and
H$\alpha$ classifications (W = weak-line, C = classical T Tauri stars) are
listed in Table 1 for reference.

\begin{table}[ht!]
\caption{H$\alpha$ emission line stars near LkH$\alpha$ 101}
\smallskip
\begin{center}
{\small
\begin{tabular}{ccccccc}
\tableline
\noalign{\smallskip}
\#\tablenotemark{a} & $\alpha$ [J2000] & $\delta$ [J2000] & $R_c$ & $K$ & H$\alpha$\tablenotemark{b} & excess\tablenotemark{c} \\
\noalign{\smallskip}
\tableline
\noalign{\smallskip}
10  & 04 29 56.35 & +35 17 43.0 & ..... & 14.31 & W & $\surd$ \\
27  & 04 29 58.61 & +35 16 17.4 & 20.99 & ..... & C & .....   \\
30  & 04 29 59.19 & +35 18 48.6 & 21.80 & ..... & C & .....   \\
32  & 04 29 59.72 & +35 13 34.3 & 15.73 & ..... & W & .....   \\
44  & 04 30 00.63 & +35 17 18.4 & 18.08 & 11.40 & W & .....   \\
63  & 04 30 02.21 & +35 17 16.8 & ..... & 16.23 & W & .....   \\
70  & 04 30 03.20 & +35 14 21.5 & 21.69 & 13.33 & W & .....   \\
72  & 04 30 03.58 & +35 16 38.0 & 18.19 & 11.86 & W & .....   \\
78  & 04 30 04.16 & +35 16 27.5 & 20.10 & 11.98 & C & .....   \\
83  & 04 30 04.62 & +35 15 01.6 & 22.08 & 13.12 & C & $\surd$ \\
95  & 04 30 05.89 & +35 17 02.7 & 19.70 & 12.28 & C & $\surd$ \\
100 & 04 30 06.65 & +35 17 53.1 & 20.19 & 12.58 & C & .....   \\
105 & 04 30 07.43 & +35 14 58.6 & 16.71 & ..... & C & .....   \\
107 & 04 30 07.50 & +35 17 54.4 & 18.32 & 11.38 & W & .....   \\
111 & 04 30 07.75 & +35 15 49.0 & 18.38 & 11.18 & C & $\surd$ \\
112 & 04 30 07.82 & +35 14 09.7 & 17.84 & 11.77 & C & .....   \\
118 & 04 30 08.36 & +35 14 39.8 & 17.63 & 10.11 & W & .....   \\
122 & 04 30 08.74 & +35 14 38.3 & 17.15 & ..... & C & .....   \\
126 & 04 30 08.97 & +35 14 33.3 & 19.51 & 10.83 & C & .....   \\
132 & 04 30 09.43 & +35 17 41.0 & 20.33 & 12.05 & C & $\surd$ \\
139 & 04 30 09.85 & +35 14 17.1 & 19.83 & 12.57 & C & $\surd$ \\
140 & 04 30 09.92 & +35 15 54.7 & 18.85 & 10.61 & C & $\surd$ \\
151 & 04 30 10.59 & +35 16 56.2 & 18.81 & ..... & C & .....   \\
157 & 04 30 11.08 & +35 16 04.0 & ..... & 11.52 & C & $\surd$ \\
180 & 04 30 13.05 & +35 13 59.5 & 12.97 & ..... & W & .....   \\
187 & 04 30 13.41 & +35 18 11.4 & 18.09 & ..... & W & .....   \\
192 & 04 30 14.26 & +35 17 51.9 & 20.20 & 12.94 & W & $\surd$ \\
194 & 04 30 14.44 & +35 16 24.5 & 13.33 & ..... & C & .....   \\
205 & 04 30 15.64 & +35 17 38.4 & 21.18 & 10.94 & C & $\surd$ \\
215 & 04 30 16.56 & +35 15 42.7 & 19.40 & 11.59 & C & .....   \\
225 & 04 30 17.24 & +35 15 38.8 & 16.10 & 10.01 & W & .....   \\
233 & 04 30 18.06 & +35 18 18.8 & 20.75 & ..... & W & .....   \\
243 & 04 30 19.35 & +35 14 00.7 & 20.81 & ..... & C & .....   \\
253 & 04 30 19.79 & +35 14 21.9 & 20.07 & 12.79 & C & .....   \\
303 & 04 30 30.41 & +35 18 34.4 & 20.19 & ..... & C & .....   \\
304 & 04 30 30.49 & +35 17 45.5 & 20.07 & 11.80 & C & .....   \\
\noalign{\smallskip}
\tableline
\noalign{\smallskip}
\multicolumn{7}{l}{\parbox{0.8\textwidth}{$^a$~Numbering system in machine-readable Table 1 of \citet{herbig04}.  LkH$\alpha$ 101 = 194.}}\\[2ex]
\multicolumn{7}{l}{\parbox{0.8\textwidth}{$^b$~H$\alpha$ emission line classification: W = weak line T Tauri star, C = classical T Tauri star.}}\\[2ex]
\multicolumn{7}{l}{\parbox{0.8\textwidth}{$^c$~A check mark notes the presence of emission in excess of the photosphere at 2.2\,$\mu$m.}}\\
\end{tabular}
}
\vspace{-0.8cm}
\end{center}
\end{table}

Classification spectra for $\sim$40 low-mass stars were compared with optical
colors to infer a mean cluster distance of 700\,pc and visual extinction of
$\sim$3.5 magnitudes.  This spectroscopic parallax distance is in agreement
with the earlier type stars in the cluster \citep[see][their
Fig.~6]{herbig04}.  Although a large spread in color-magnitude space exists
within the cluster, the H$\alpha$ emission line stars (marked with crosses in
Fig.~3) have a median age around 0.5 Myr using the aforementioned properties
and isochrones.

\begin{figure}[ptb]
\begin{center}
\fbox{
\includegraphics[height=2.3in,draft=False]{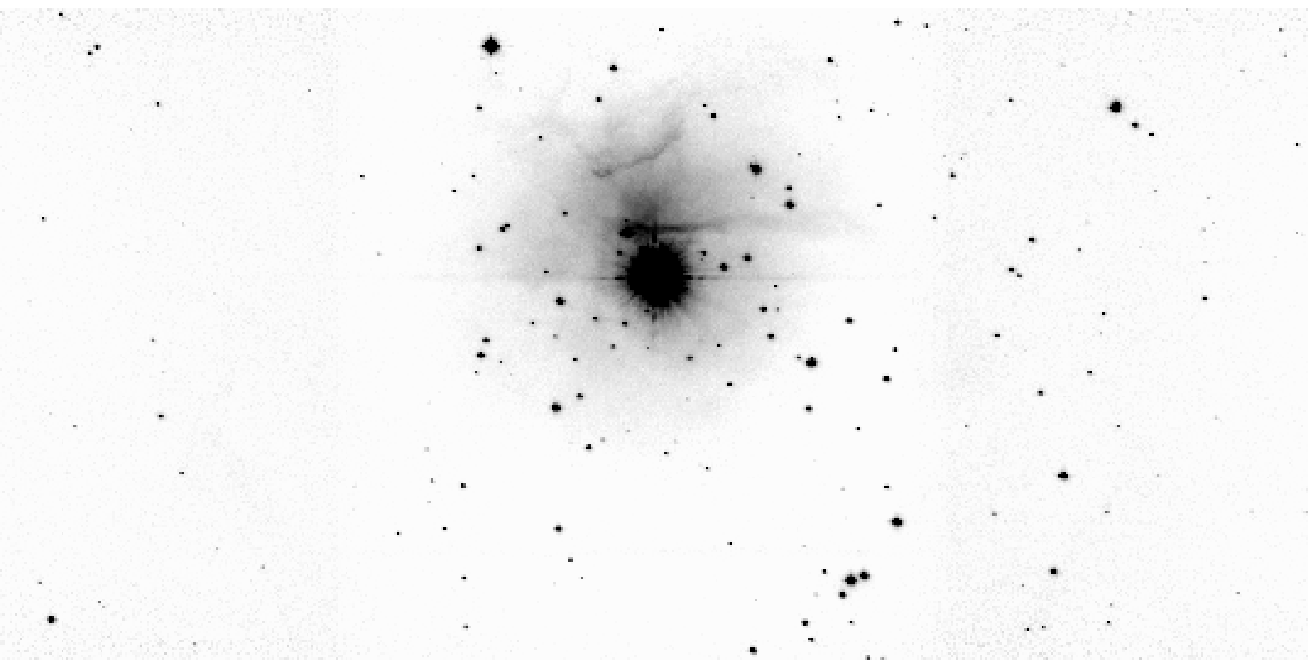}
}
\caption{$K$-band image of the LkH$\alpha$ 101 cluster, covering 8\arcmin\ E-W
and 4\arcmin\ N-S \citep[from][]{herbig04}.  The young star cluster is more
apparent than in the optical (see Fig.~1), along with some interesting nebular
features near the central source.}
\vspace{1.2cm}
\includegraphics[height=3.25in,draft=False]{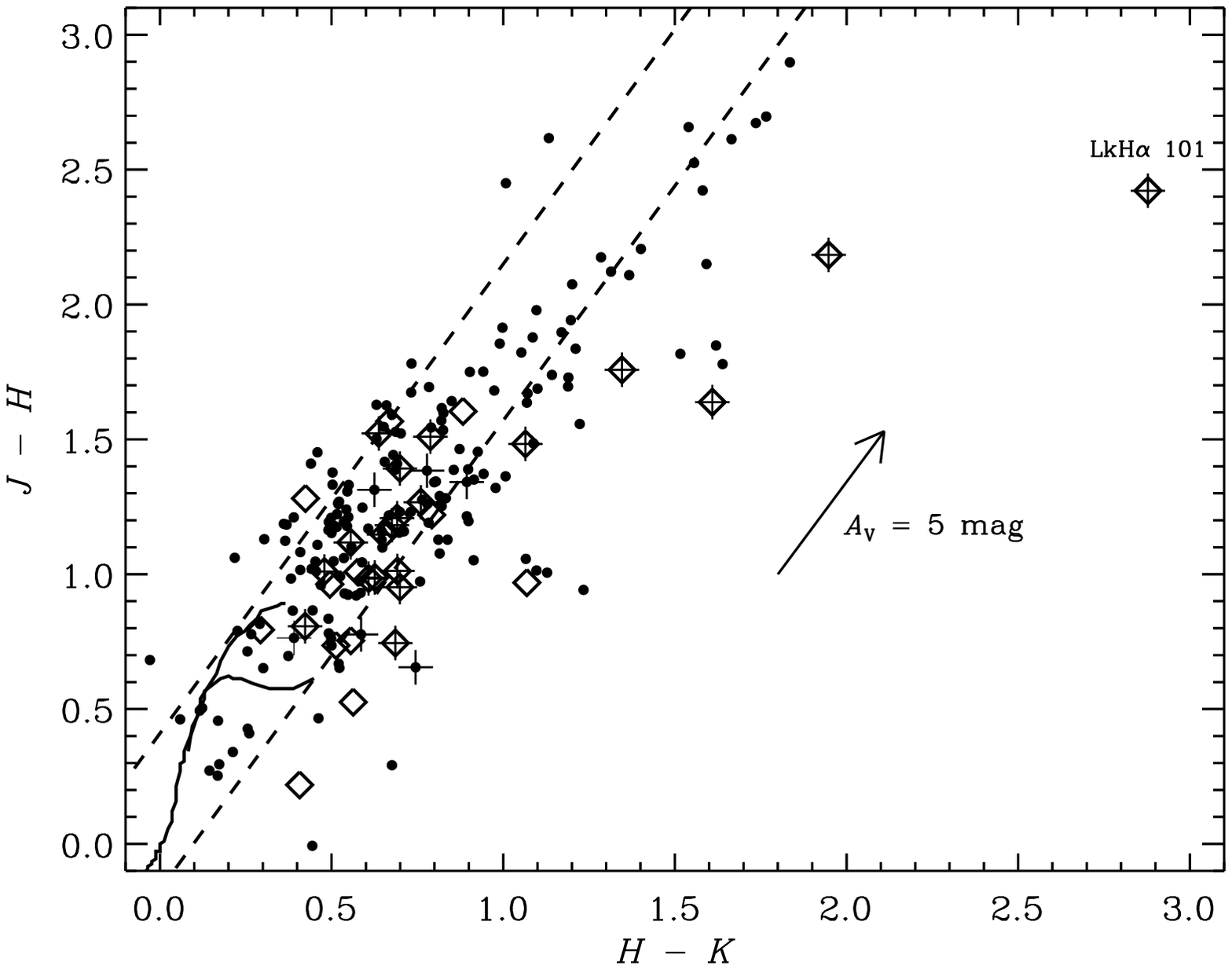}
\caption{Near-infrared $J-H$, $H-K$ color-color diagram for the LkH$\alpha$ 101
cluster \citep[from][]{herbig04}.  The symbols are as in Figure 3.  Solid
curves mark normal main sequence and giant colors, and dashed lines define the
reddening band.}
\end{center}
\end{figure}

In addition to these optical data, \citet{herbig04} obtained a deep ($K \le
18.5$) set of $JHK$ images of the region in an effort to search for the
near-infrared excesses characteristic of the warm inner regions of
circumstellar disks.  A $K$-band mosaic image of the region is shown in Figure
4.  Using the near-infrared $J-H$, $H-K$ color-color diagram exhibited in
Figure 5, these authors inferred that $\sim$60 sources had an excess at
2.2\,$\mu$m; most of those are listed separately in Table 2, along with their
positions and $K$-band magnitudes.  \citet{haisch00} have pointed out, however,
that excesses determined for such short wavelengths may not be representative
of inner disk emission for a number of reasons: of particular concern here is
the bright infrared nebulosity.  Of these excess stars, 8 have $J-K > 4$ and 18
others have $3 \le J-K \le 4$.  This is in general agreement with the
suggestions of \citet{barsony91} and \citet{aspin94} that an even younger, more
embedded, population of stars may exist in the material surrounding LkH$\alpha$
101.

\begin{table}[ht!]
\caption{Near-infrared excess sources surrounding LkH$\alpha$ 101}
\smallskip
\begin{center}
{\small
\begin{tabular}{cccc|cccc}
\tableline
\noalign{\smallskip}
\#\tablenotemark{a} & $\alpha$ [J2000] & $\delta$ [J2000] & $K$ & \#\tablenotemark{a} & $\alpha$ [J2000] & $\delta$ [J2000] & $K$ \\
\noalign{\smallskip}
\tableline
\noalign{\smallskip}
6   & 04 29 55.82 & +35 16 40.0 & 15.07 & 184 & 04 30 13.17 & +35 16 33.6 & 15.32 \\
22  & 04 29 58.25 & +35 15 35.3 & 14.68 & 188 & 04 30 13.44 & +35 15 41.5 & 14.98 \\
34  & 04 29 59.94 & +35 15 15.3 & 14.62 & 197 & 04 30 15.17 & +35 15 30.6 & 15.24 \\
46  & 04 30 00.76 & +35 17 57.7 & 14.33 & 198 & 04 30 15.20 & +35 16 40.4 & 10.17 \\
50  & 04 30 01.24 & +35 14 29.2 & 14.41 & 200 & 04 30 15.27 & +35 16 33.3 & 12.28 \\
82  & 04 30 04.59 & +35 16 04.4 & 12.45 & 211 & 04 30 16.11 & +35 16 10.0 & 12.50 \\
89  & 04 30 05.52 & +35 17 08.2 & 14.76 & 212 & 04 30 16.30 & +35 15 24.7 & 11.86 \\
117 & 04 30 08.24 & +35 14 10.7 & 13.33 & 214 & 04 30 16.46 & +35 14 38.9 & 14.40 \\
133 & 04 30 09.51 & +35 14 41.1 & 12.95 & 222 & 04 30 17.13 & +35 16 16.3 & 10.38 \\
141 & 04 30 09.97 & +35 15 38.4 & 11.44 & 226 & 04 30 17.25 & +35 16 03.8 & 13.20 \\
150 & 04 30 10.57 & +35 16 50.3 & 10.21 & 228 & 04 30 17.37 & +35 15 21.4 & 15.93 \\
154 & 04 30 10.89 & +35 16 13.3 & 13.93 & 229 & 04 30 17.55 & +35 16 26.6 & 13.31 \\
155 & 04 30 10.94 & +35 16 21.3 & 14.21 & 230 & 04 30 17.74 & +35 17 13.7 & 13.38 \\
167 & 04 30 11.76 & +35 16 31.7 & 11.00 & 232 & 04 30 17.92 & +35 16 08.4 & 13.24 \\
168 & 04 30 12.19 & +35 14 51.0 & 13.77 & 236 & 04 30 18.68 & +35 16 42.9 & 12.46 \\
170 & 04 30 12.23 & +35 15 47.3 & 12.65 & 237 & 04 30 18.80 & +35 16 41.9 & 11.88 \\
173 & 04 30 12.34 & +35 16 28.4 & 10.70 & 244 & 04 30 19.39 & +35 15 57.3 & 10.92 \\
175 & 04 30 12.77 & +35 17 21.3 & 11.72 & 247 & 04 30 19.46 & +35 16 34.9 & 11.80 \\
178 & 04 30 13.01 & +35 16 33.3 & 12.66 & 284 & 04 30 26.96 & +35 14 49.0 & 15.53 \\
181 & 04 30 13.08 & +35 15 18.8 & 15.52 & 288 & 04 30 27.99 & +35 15 15.7 & 13.80 \\
182 & 04 30 13.09 & +35 16 31.2 & 14.80 & 292 & 04 30 28.54 & +35 15 51.1 & 15.71 \\
\noalign{\smallskip}
\tableline
\multicolumn{8}{l}{\parbox{0.8\textwidth}{\footnotesize $^a$ Numbering system of \citet{herbig04}.}}
\end{tabular}
}
\end{center}
\end{table}

\section{LkH$\alpha$ 101}

Since Herbig's (1956) identification of LkH$\alpha$ 101 as the illuminating
source of the NGC 1579 reflection nebula, this still-enigmatic object has
become one of the most thoroughly studied young stars in the sky.  Early
spectroscopic observations identified a remarkably strong H$\alpha$ emission
line (equivalent width of $\sim$550\,\AA; Herbig et al.~2004) and a series of
other atomic emission features, dominated by permitted and forbidden lines of
singly-ionized iron \citep{herbig56,herbig71,allen73,thompson76}.  Assuming $d
= 700$\,pc, the position of LkH$\alpha$ 101 in an H-R diagram is consistent
with an early B star (B0 or B1) on or near the main-sequence with a visual
extinction of roughly 10 magnitudes \citep{herbig04}.  Extinction estimates for
the source vary significantly, but $A_V \approx 10$ lies comfortably in the
center of the range of values.  This spectral classification is in good
agreement with that implied by the radio continuum observations, which require
a Lyman continuum flux from a $\sim$B0.5 main-sequence star to explain the
observed H\,II region emission
\citep{harris76,brown76,becker88,hoare94,hoare95}.  Despite these hints at the
underlying radiation source, no stellar absorption features have ever been
clearly seen for LkH$\alpha$ 101 \citep{herbig04}.  There is no \emph{direct}
spectroscopic evidence with which to classify the LkH$\alpha$ 101 photosphere.

\begin{figure}
\begin{center}
\includegraphics[height=2.5in,draft=False]{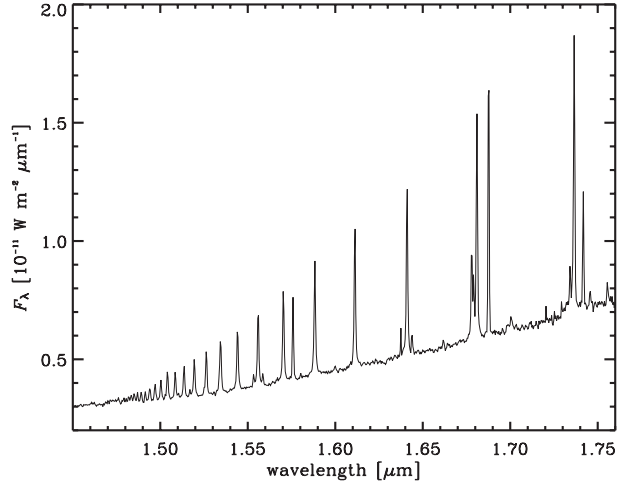}
\end{center}
\caption{Portion of the near-infrared spectrum of LkH$\alpha$ 101 (taken with
the SpeX instrument on the 3.0-m IRTF telescope).  Brackett series lines can be
seen to Br\,(42), in addition to some metal lines and a red continuum.}
\end{figure}

Many of the spectroscopic studies of LkH$\alpha$ 101 have focused on
understanding the physical conditions in the region(s) where the emission line
spectrum is generated.  In the near-infrared (1-5\,$\mu$m), the spectrum is
dominated by H\,I lines in the Paschen and Brackett series, along with various
transitions of He\,I, Fe\,II, O\,I, and Mg\,II, among others
\citep{thompson76,thompson76b,thompson77,simon84,hamann89,rudy91}.  The optical
spectrum has similar contributors \citep{hamann89,kelly94,herbig04}.  The
oxygen and magnesium lines are thought to be excited by Bowen fluorescence from
Ly\,$\beta$ photons from the hot star \citep{hamann89}.  From this and the
presence of high Paschen and Balmer series lines, some of which are shown in
Figure 6, it is clear that the emission-line spectrum is at least partially
generated in a high density, circumstellar environment.

\begin{figure}[ht!]
\begin{center}
\includegraphics[height=2.5in,draft=False]{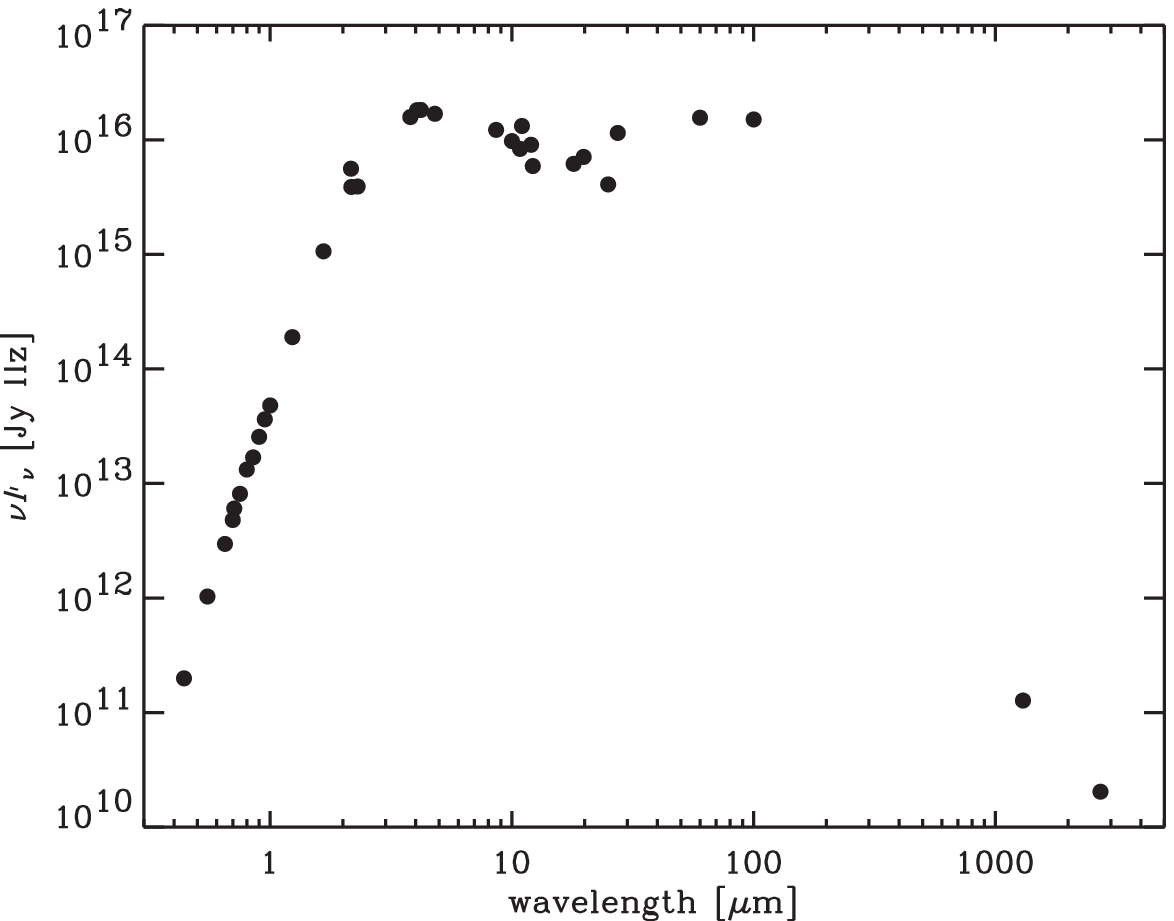}
\end{center}
\caption{The broadband spectral energy distribution of LkH$\alpha$ 101,
compiled from the literature \citep{cohen71,price83,simon84,barsony90,weaver92,osterloh95,danen95,cutri03,herbig04}.}
\end{figure}

Similar conclusions are reached based on forbidden emission line ratios
(Kelly et al.~1994; Herbig et al.~2004) and the slope of the radio continuum
\citep{brown76}.  \citet{hamann89} and \citet{herbig04} explored the
possibility that electron scattering in such a dense circumstellar environment
could broaden the standard early-type photospheric absorption lines into the
continuum.  The latter authors have ruled this out, and so the absence of these
lines remains an unresolved issue.  However, for photons to escape and produce
the observed H\,II region, the star cannot be completely enveloped in such
high-density material; the circumstellar emission line region must be
geometrically anisotropic \citep[e.g.,][]{simon84,hamann89}.  These
spectroscopic properties have led to comparisons of LkH$\alpha$ 101 and evolved
massive stars which have moved off the main-sequence \citep[e.g., $\eta$ Car,
MWC 300, MWC 349;][]{herbig71,allen73,thompson76,hamann89}.  While
acknowledging the very different evolutionary states of these objects and
LkH$\alpha$ 101, the physical structures responsible for their similar spectra
are likely the same: stellar winds and the irradiation of a dense circumstellar
disk.

The broadband spectral energy distribution (SED) of LkH$\alpha$ 101, displayed
in Figure 7, shows a large infrared excess indicative of thermal continuum
emission from circumstellar dust.  \citet{danen95} argued that the dip in the
SED between $\sim$10 and 20\,$\mu$m is not easily explained by the standard
circumstellar disk models \citep[e.g.,][]{adams87,beckwith90}.  However, this
is not likely a significant problem, as the large beam sizes for
long-wavelength data often result in flux overestimates from extended emission
or the excess emission from other nearby young stars in the cluster.
Regardless, the 10\,$\mu$m observations of Danen et
al.~indicated a very small emission source size ($\sim$50\,mas) and warm
characteristic dust temperature ($\sim$1000\,K).  Using the OVRO
interferometer, \citet{barsony90} detected unresolved thermal continuum
emission ($F_{\nu} \approx 185$\,mJy) at 3\,mm from this circumstellar dust.
With the standard optically thin, isothermal dust assumptions and opacity law
\citep[e.g.,][]{andrews05}, the corresponding mass of circumstellar material
(gas and dust) is estimated to be $\sim$1-2\,M$_{\odot}$, or roughly 10\%\ of
the proposed stellar mass.

\begin{figure}[ht!]
\plotfiddle{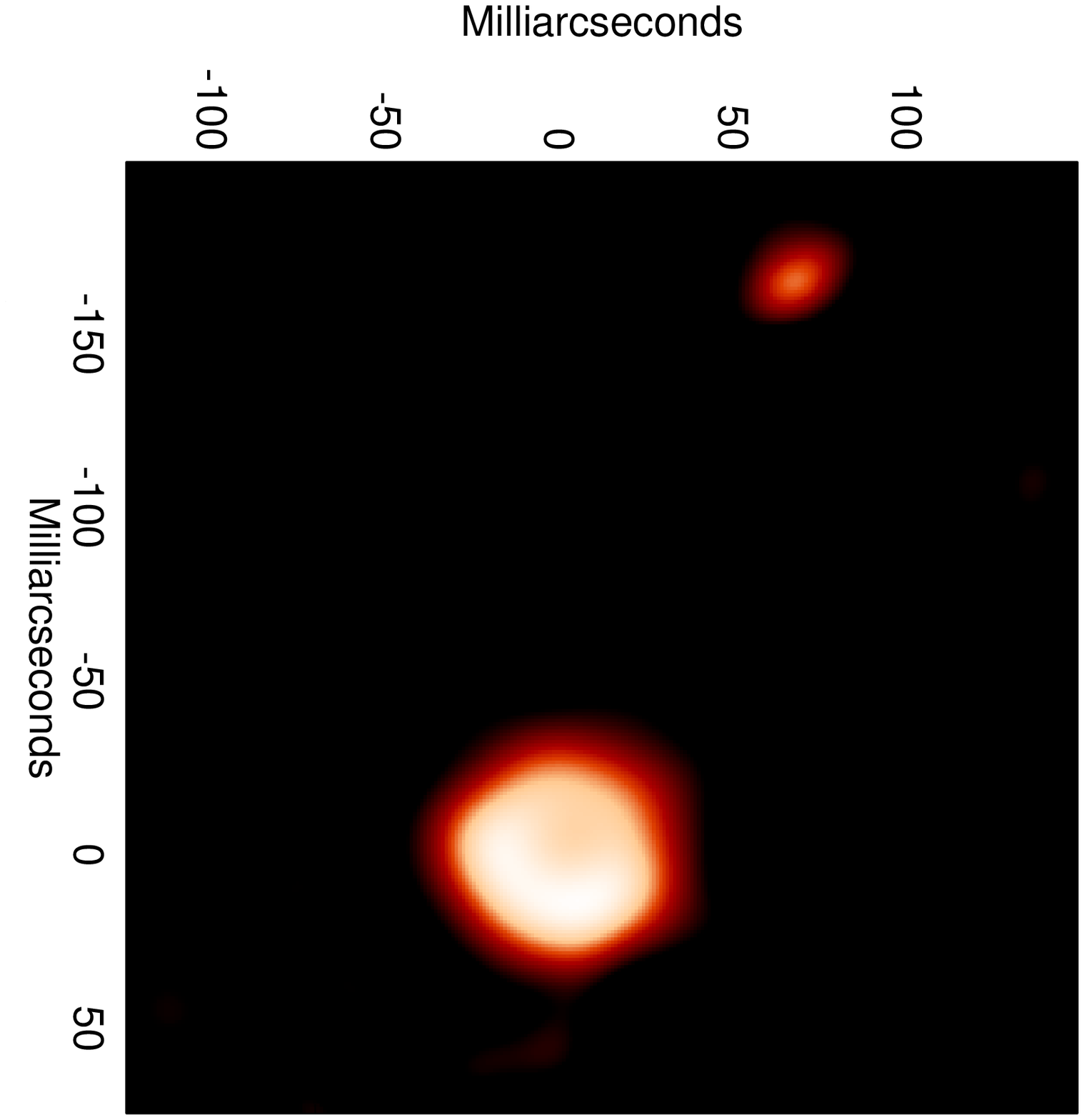}{1.9in}{90}{35}{35}{60}{-40}
\plotfiddle{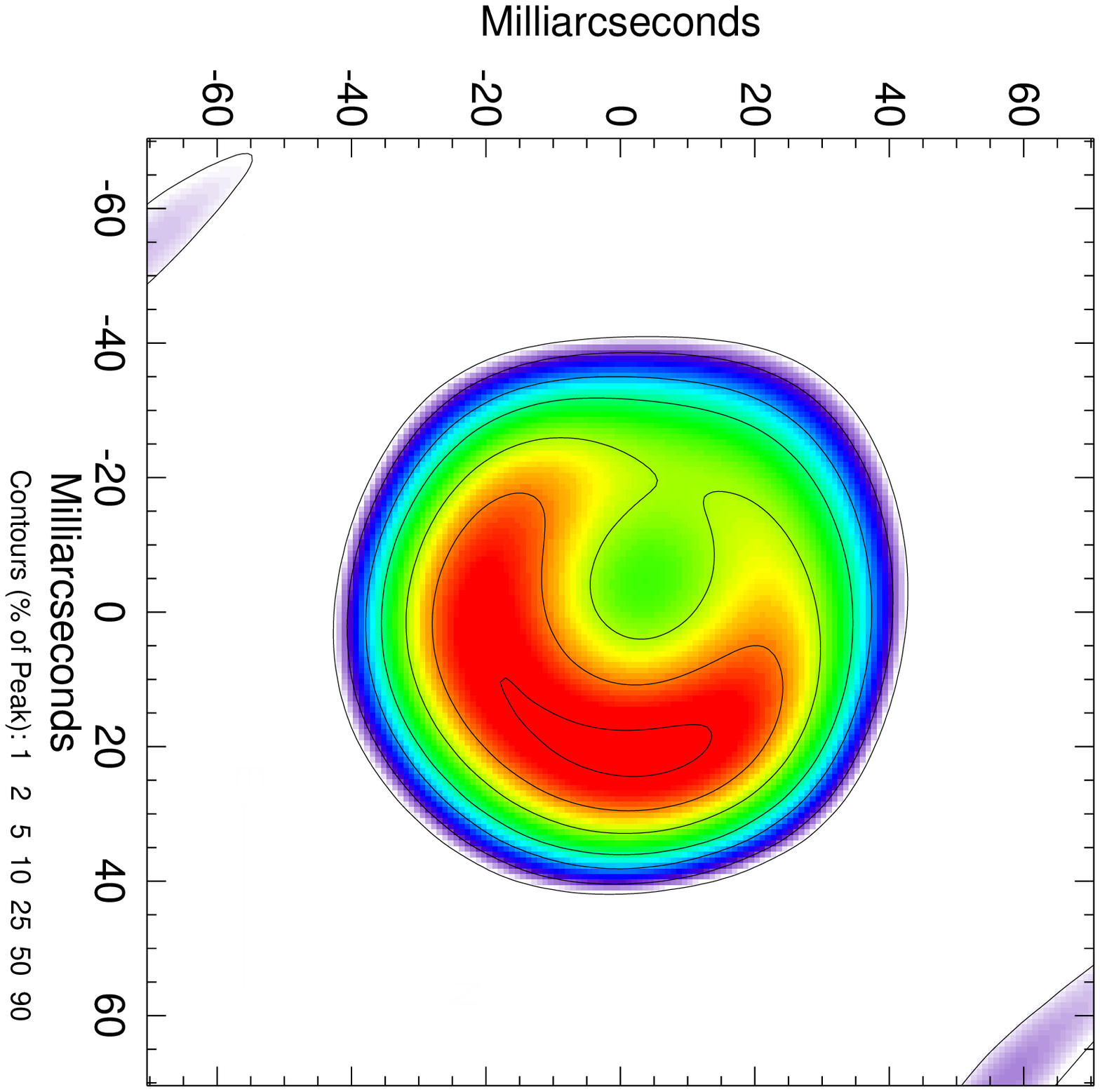}{0.0in}{90}{35}{35}{250}{-15}
\caption{({\it left}) High-resolution $H$-band image showing the bright disk
around LkH$\alpha$ 101 and a faint companion star off to the northeast.  ({\it
right}) Detailed $K$-band image of the LkH$\alpha$ 101 disk, with a depressed
central cavity and nearly face-on orientation.  Images adapted from
\citet{tuthill01,tuthill02}, courtesy of P.~Tuthill.}
\end{figure}

The remarkably bright infrared emission from LkH$\alpha$ 101 ($K \approx 3$)
made it an ideal test subject for the rapidly developing technologies involved 
in very high-resolution infrared imaging.  In a pioneering study by 
\citet*{tuthill01}, high angular resolution images showed that the infrared 
emission originatesin a nearly face-on, resolved (FWHM = 40\,mas) disk 
structure with a large central cavity surrounding the star.  These and other 
data were shown to be consistent with high-mass disk models that call for an 
inner region cleared by the sublimation of dust particles by high-energy 
stellar irradiation and a thick, flared geometry \citep{tuthill01,tuthill02}.  
Moreover, these same data showed that the infrared emission morphology of the 
disk actually changes with time and revealed the presence (and relative proper 
motion) of a faint, blue companion star $\sim$0\farcs2 to the northeast.  
Figure 8 shows a wider field $H-$band image with the LkH$\alpha$ 101 disk and 
this companion star, as well as a more detailed look at the disk morphology.  
\citet{tuthill02} used the motion of the companion star and properties of 
LkH$\alpha$ 101 and its disk to estimate an intermediate distance to the 
source, $d \approx 340$\,pc. \citet{herbig04} noted that the disk geometry 
proposed by Tuthill et al.~could account for the observed splitting of the 
optical Fe\,II lines, offering an interesting connection between the gas and 
dust in the inner disk.

\section{Recent Results and Recommended Future Work}

The embedded cluster around LkH$\alpha$ 101 has been the object of much recent
scrutiny, as the subject of an infrared survey with the {\it Spitzer Space
Telescope} and a simultaneous campaign with the {\it Chandra X-ray
Observatory} and the VLA \citep{wolk08,osten08}.  Images from the {\it Spitzer}
and {\it Chandra} observations are shown together in Figure 9.  The
mid-infrared {\it Spitzer} photometry reveals the presence of 16 protostars
(Class I sources) and an additional 95 T Tauri stars (Class II sources), along
with 9 ``transition" objects that show large 24\,$\mu$m excesses but only
photospheric emission at shorter wavelengths.  The latter are widely
interpreted to be circumstellar disks with evacuated inner regions.  The {\it
Chandra} observations identify an additional 65 X-ray sources coincident with
infrared stars with near-photospheric colors, consistent with their association
as more evolved (Class III) cluster members.  This brings the total list of
known cluster membership to $\sim$185, many of which are actively being
confirmed spectroscopically \citep{winston08}.

\begin{figure}[p]
\begin{center}
\includegraphics[width=0.9\textwidth,draft=False]{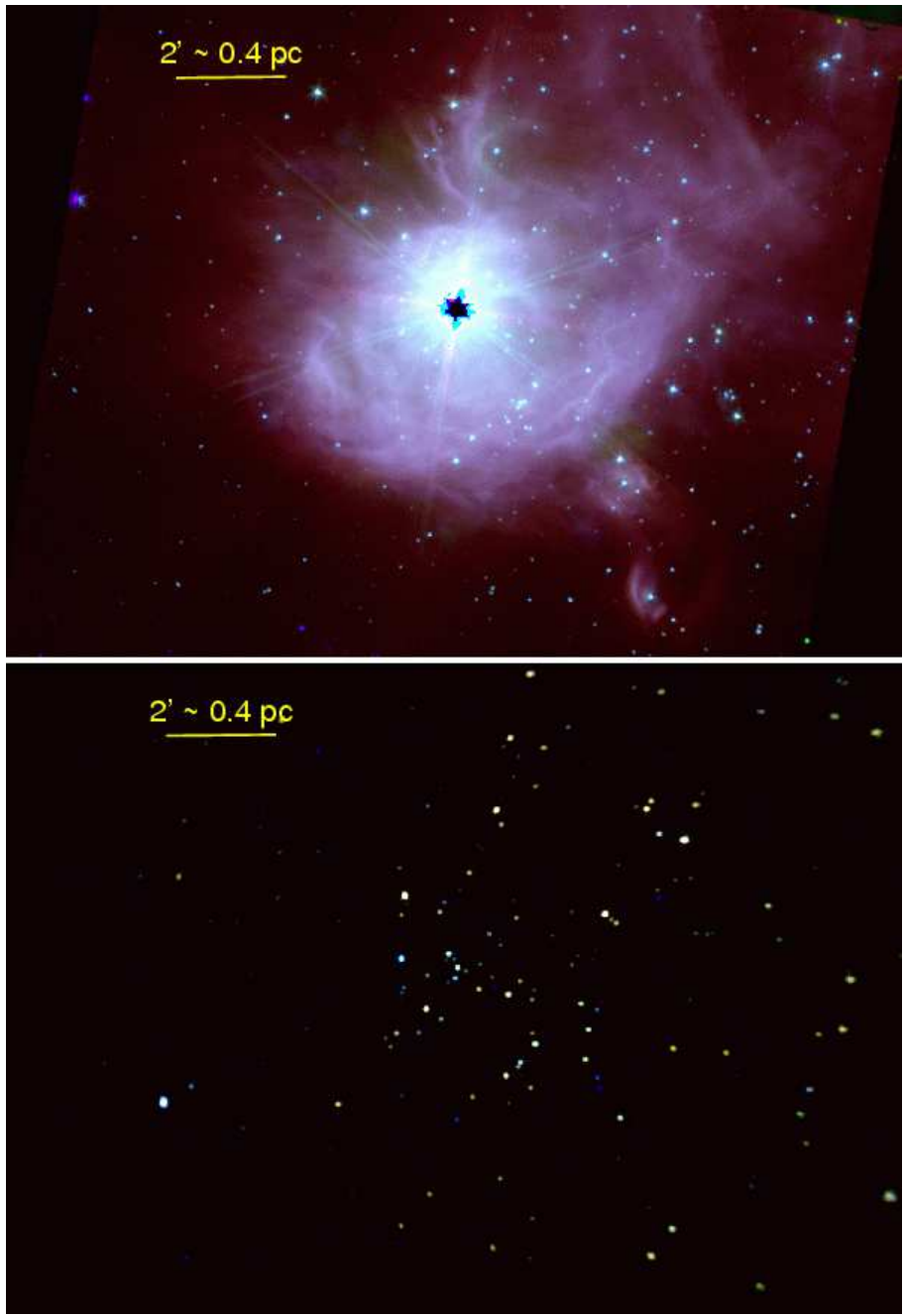}
\end{center}
\caption{False color images of the LkH$\alpha$ 101 cluster at {\it Spitzer}
mid-infrared (top) and {\it Chandra} X-ray (bottom) wavelengths
\citep{wolk08}.  The infrared image is a composite of data at 3.6 (blue), 4.5
(green), and 5.8\,$\mu$m (red) on a logarithmic scale.  The X-ray image is a
composite of three energy ranges: 0.5-1.5\,keV (red), 1.1-2.4\,keV (green), and
2.1-8.0\,keV (blue).}
\end{figure}

In addition to identifying new members and characterizing their evolutionary
states via their infrared excess properties, these new data can be utilized to
estimate the total cluster size in two complementary ways.  The first method
exploits the empirical similarity of the X-ray luminosities from Class II and
III sources \citep[][and references therein]{feigelson99}.  Using this
assumption that the same fraction of Class II and III infrared sources should
be detected with X-rays, the new observations imply that the $\sim$45\%\ X-ray
detection rate of Class II sources would translate to $\sim$82 undetected Class III sources.  This estimate would bring the total cluster membership to
$\sim$270 sources.  The second method relies on the derived shape of the
universal X-ray luminosity function (XLF) for young clusters asserted by
\citet{feigelson05}.  If this universal XLF applies for the LkH$\alpha$ 101
cluster, the total cluster membership should be in the range of 280-330 stars,
and the smaller cluster distances are firmly ruled out in favor of values in
the range $d \approx 550$-750\,pc.  The closer end of this range brings the two
methods of estimating cluster membership numbers into good agreement.

This recent work is encouraging in its focus on establishing a firm cluster
membership base for future work.  Only when a relatively complete membership
roster has been obtained can a more comprehensive analysis of the distance,
age, initial mass function, circumstellar disk fraction, and other basic
properties be derived.  Although of less immediate importance, a closer
examination of individual cluster members would certainly be interesting.
\citet{herbig04} present a curious high-resolution spectrum of the bright,
nebulous star immediately to the northeast of LkH$\alpha$ 101 (their ``Star D",
HBC 391; see Fig.~1), showing it to be an early K giant with a number of
strange features.  Those authors also note that none of the 5 B-type stars
apparently associated with the same dark cloud (excluding LkH$\alpha$ 101) show
the standard signatures of youth noted for other Herbig Be stars.

Given the high concentration of infrared sources in the immediate vicinity of
LkH$\alpha$ 101 noted by \citet{herbig04}, high angular resolution infrared
images using adaptive optics would provide an interesting complement to the new
{\it Spitzer} data.  Some of the most interesting cluster members may be
lurking in the tremendous glare of the central source, including the small
group of stars identified by \citet{herbig04} that lie at one end of a ``bar"
of infrared nebulosity (see Fig.~4).  Along these same lines, those interested
in the formation and evolution of massive stars and their disks should make a
concerted effort to follow up the high-resolution infrared work of Tuthill and
colleagues.  High resolution (sub-arcsecond) millimeter observations of the
dust continuum and various molecular line transitions would aid tremendously in
interpreting the circumstellar environment around LkH$\alpha$ 101.  As with
many other clusters in this book, the LkH$\alpha$ 101 region still holds a lot
of promise for future observations with more sensitive, higher resolution
instrumentation. \\

\acknowledgments{S.A.~is very grateful to George Herbig and Scott Dahm for
useful conversations and advice.  We would like to thank Peter Tuthill for
kindly providing the images in Figure 8.}

\end{document}